# Optimization of Surface Plasmon Resonance Biosensor for Analysis of Lipid Molecules

Ehsan Kabir[a], Syed Mohammad Ashab Uddin[b] and Sayeed Shafayet Chowdhury[c,*]

[a]Bangladesh University of Engineering and Technology, Dhaka 1205, Bangladesh, [b]University of California, Irvine, CA 92697, USA, [c]Purdue University, Indiana, USA, *email: chowdh23@purdue.edu

## Abstract

Surface Plasmon Resonance (SPR) is an important bio-sensing technique for real-time label-free detection. However, it is pivotal to optimize various parameters of the sensor configuration for efficient and highly sensitive sensing. To that effect, we focus on optimizing two different SPR structures- the basic Kretschmann configuration and narrow groove grating. Our analysis aims to detect two different types of lipids known as phospholipid and eggyolk, which are used as analyte (sensing layer) and two different types of proteins namely tryptophan and bovine serum albumin (BSA) are used as ligand (binding site). For both the configurations, we investigate all possible lipid-protein combinations to understand the effect of various parameters on sensitivity, minimum reflectivity and full width half maximum (FWHM). Lipids are the structural building block of cell membranes and mutation of these layers by virus and bacteria is one the prime reasons of many diseases in our body. Hence, improving the performance of a SPR sensor to detect very small change in lipid holds immense significance. We use finite-difference time-domain (FDTD) technique to perform quantitative analysis to get an optimized structure. We find that sensitivity increases when lipid concentration is increased and it is the highest (21.95°/RIU) for phospholipid and tryptophan combination when metal and lipid layer thickness are 45 nm and 30 nm respectively. However, metal layer thickness does not cause any significant variation in sensitivity, but as it increases to 50 nm, minimum reflectivity and full width half maximum (FWHM) decreases to the lowest. In case of narrow groove grating structure, broad range of wavelengths can generate SPR and the sensitivity is highest (900nm/RIU) for a configuration of 10 nm groove width and 70 nm groove height at a resonance wavelength of 1411 nm.

Keywords : Biosensor, SPR, FDTD, FWHM, FOM, Grating, Kretschmann, Lipid, Protein, Plasmon, Refractive index, Sensitivity.

## Introduction

In recent years, various biosensors have been widely used for improvement of our lives ranging from heart-rate measurement [1], [2], remote healthcare [3]–[5] to personalized wheelchair control [6]–[9]. Among several biosensing techniques, optical sensors are extremely popular for accurate and fast operation. Surface plasmon resonance (SPR) sensors are one such important optical instruments with many applications including study of biomolecular interactions and chemical reactions, monitoring environmental pollution, detection and diagnosis of diseases and examination of new drugs [10], [11]. It is a label-free, cheap and simple detection technique providing real time data in the investigation of medicine, chemicals and biomolecules. Environmental conditions such as temperature, pressure, humidity can cause the change in refractive index which is easily detectable by SPR systems. Chemical or physical adsorption also changes the refractive index of the surface of SPR sensors. Moreover, biological events like protein-lipid, protein-DNA, antigen-antibody interactions can be detected by SPR. The experiments with biosensors involve immobilizing one reactant on a surface and monitoring its interaction with another reactant in solution [12], [13]. We can measure the kinetics of interactions and quantify the strength of bonds among various analytes by measuring their physical properties such as mass, volume, refractive index, dielectric permittivity etc. by SPR biosensor. The interaction events are captured by measuring the changes in these physical properties at the surface using a transducer which transforms the changes in the physical properties into a measurable signal (e.g. current or voltage) for recording in real-time.

The lipid bilayer structure of cellular membrane has severe impact on the transport of drugs through the membrane or on the entrance of virus and bateria to cell [14]. It can be affected by many diseases. For example, the toxicity of amyloid $\beta$-peptide ($A\beta$) responsible for Alzheimer's disease intensively perturbs the stability of lipid bilayers of cell membranes [15]. $A\beta 42$ peptides decrease the effective width of the hydrophobic region. Besides, lipid bilayer in the red blood cell is affected by hemoglobiunopathy diseases. Fats which are sources of energy also contain lipids. However, abnormalities in specific enzymes responsible for breaking down fats enforce accumulation of lipids causing various diseases [16]. The different states such as gas phase, liquid phase, solid phase, mixed phase of the diseased lipid bilayers can be identified by the change in the refractive index of the lipid bilayers. Many experimental works have been done for finding lipid-protein interactions in real time using SPR techniques by immobilizing lipid on sensor chip and passing protein solution over it [17]. Some simulation works have been done with either lipid or protein as the sensing layer to find out how the shift of resonance angle varies with the thickness of lipid or protein layer or how sensitivity differs in different configurations of SPR sensors [18], [19]. Other simulation based work showed how to increase sensitivity and other performance parameters by adding different layers to the basic SPR configuration [20], [21]. However, to extract the utmost utility out of any sensor configuration, it is crucial to examine the effect of various design choices on metrics of interest which motivates our work. In order to thoroughly analyze the SPR configurations in terms of the design parameters, we perform a comprehensive study in this paper.

We use combinations of lipid and protein layers in the basic Kretschmann configuration and grating configuration and change structural parameters like thickness, height and width and optical parameters like refractive index to figure out how sensitivity, FWHM, shift in resonance angle or FOM are changed and which of the simulated configuration provides best performance under each condition. Our study will be helpful for future experimental work by providing the optimum values for various design parameters.

## SIMULATION METHODOLOGY

The Finite-Difference Time-Domain method (FDTD) is the most modern technique for solving Maxwell's equations to analyze electromagnetic waves in a simple way [22]. It gives both time domain and frequency domain information for problems related to electromagnetics. Hence, we used FDTD solution method in our work. We used Gold (Au-Palik) as metal layer and SiO2 (Glass- Palik) as substrate layer for kretschmann configuration. The substrate layer is chosen as infinitely long. No glass prism was used for grating coupler and there was no restriction for the thickness of metal layer in this configuration. For 2D simulation, the structure we used has similarities along Y axis but contains different layers along X axis. To model 632.8 nm He-Ne laser, we used plane wave source. Two boundary conditions-bloch boundary conditions for periodicity in perpendicular direction and perfectly matched layer (PML) for preventing reflection from the boundary in horizontal direction were used to maintain periodicity of our structure and equal phase and amplitude of incident light.

## SIMULATED STRUCTURE

We simulated the structure shown in Fig. 1. Amino acid named tryptophan having refractive index of 1.754 [23] and BSA with refractive index 1.6 [23] are used as ligand because Lipid can bind well with proteins and amino acids [24]. Pure Phospholipid having refractive index of 1.46 [23] is used to demonstrate the analyte of the flow cell in actual biosensor. Since flow cell has a flow rate, binding events occur at a certain rate. The refractive index changes with the bound mass of analyte. The variation of the thickness of lipid layer in simulation is equivalent to the amount of lipid that has been bound with the ligand and the concentration of lipid.

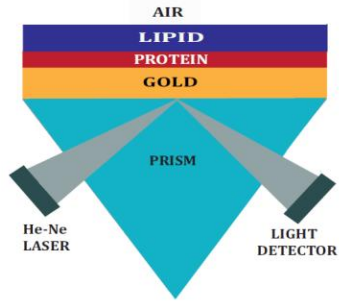

Fig. 1. Krestchmann Configuration For Simulation

The thickness of Gold is kept constant at 50nm [25] first and tryptophan layer having thickness of 1.3nm [26] is deposited on the Gold. Change in concentration of Lipid is simulated by varying its thickness. As the concentration is increased the effective refractive index [27] will increase causing greater shift of resonance angle which gives higher response unit with respect to time. The Gold layer thickness is also varied keeping the Lipid layer thickness constant. We generated the reflectance curves with respect to angle of incidence as shown in Fig. 2 demonstrating shifts in resonance angle and change in width of the curves with respect to thickness.

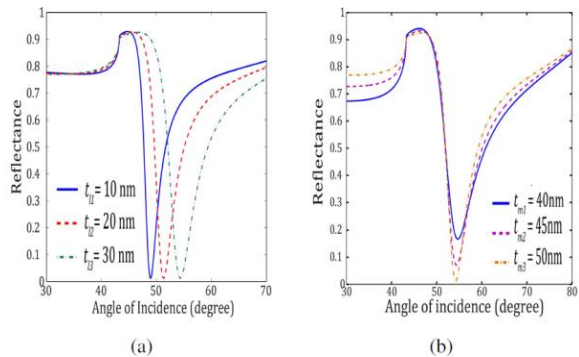

Fig. 2. Reflectance vs Angle of incidence, (a) for different lipid layer thickness keeping the metal layer fixed at 50nm, (b) for different metal layer thickness keeping the lipid layer fixed at 30nm

Tryptophan was substituted with BSA and the same procedures were executed. We also substituted Gold with Silver. Then we replaced phospholipid by eggyolk which is another lipid found in fats. The full width half maximum (FWHM), sensitivity, minimum reflectance and figure of merit were calculated to find out the optimized structure. It is easier to sense the presence of any analyte if full width half maximum is small meaning a sharp change of angle which can be distinguished precisely. Minimizing minimum reflectivity is also necessary because it indicates higher coupling of light with plasmons which makes detection easier. To combine all the performance parameters, the figure of merit (FOM) which is equal to Sensitivity/(FWHM*Minimum Reflectance) is calculated.

# RESULTS AND DISCUSSIONS

We generated reflectance curves and electric field profiles from which various performance parameters were measured and compared using graphs and tables such as Table I, Table II and Table III for different thickness of metal and lipid and for different combinations of lipid and protein. Sensitivity analysis was done by varying both the angle of incidence and the wavelength of light.

*A. Angular Interrogation*

The most vital parameter of a sensor is sensitivity which is defined in this case as the angular shift with respect to the variation of refractive index. From our data we don't observe any significant effect on sensitivity with the metal layer thickness variations. In addition, resonance angle is also nearly constant. However, sensitivity is found to be proportional to lipid layer thickness meaning higher sensitivity for higher lipid concentration which is indicated by larger lipid thickness. Minimum reflectivity also increases with the increase of lipid concentration. It is due to the increment in effective refractive index which causes the reflectivity to increase although the variation is not significant. On the other hand, minimum reflectivity is inversely proportional to the metal layer thickness, and this variation is significant as shown in Fig. 3. That means the absorption coefficient of metal decreases when thickness increases for particular wavelengths. Another important parameter is FWHM which doesn't show any significant dependency on the thickness of the metal layer as seen from Fig. 4. Highest FWHM is found for 45nm of Gold, and it is lowest around 50nm. However, it has unfavorable relationship with the lipid of higher concentration. FWHM increases with the increase in concentration of Lipid. It implies that absorption coefficient is increased. Any damage of lipid will lead to a variation in refractive index which can be detected well by any of these structures as sensitivity variations are not significant. However, using tryptophan as ligand gives higher sensitivity ($21.95^0$/RIU) than that for BSA ($17.24^0$/RIU) as shown in the Tables. On the other hand, 50 nm Gold is optimum for minimum reflectivity and FWHM.

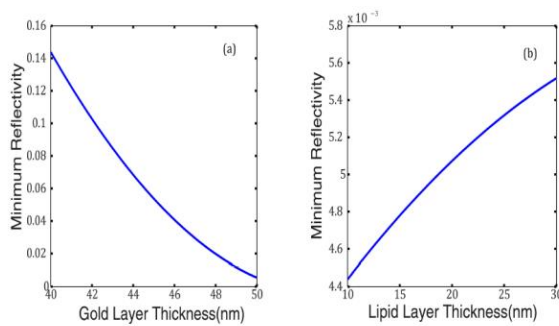

Fig. 3. Minimum Reflectivity decreasing with respect to the increase of Gold Thickness($t_m$) and showing significant reduction of minimum reflectivity

Lower minimum reflectivity and FWHM indicates high absorption at resonance angle which can be easily measured. Moreover, sensitivity is always higher for 30 nm lipid thickness. The highest FOM (522 from Table I) observed is for tryptophan and phospholipid combination at metal and lipid thickness of 50 nm and 30 nm respectively.

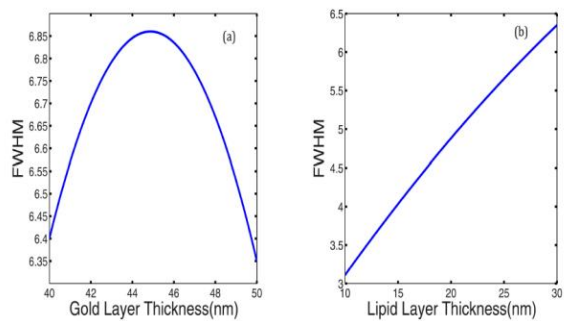

Fig. 4. (a) FWHM increases first and then decreases with a peak value for 45 nm Gold when metal thickness is varied, (b) FWHM increases with Lipid thickness

*B. Wavelength interrogation*

We studied characteristics of Lipids by wavelength interrogation. Broadband planewave source can be used for studying the behaviour at different wavelength. By varying the wavelength of our source, we found that sensitivity is lower for wavelength interrogation than angular interrogation. It means slight change in refractive index cannot be detected by this technique as can be seen by very small shift in the reflectance curve of Fig. 5.

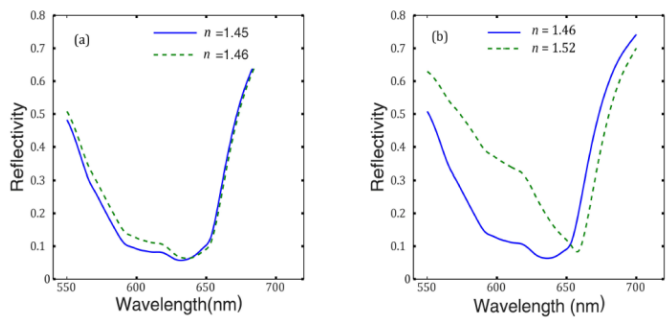

Fig. 5. Sensitivity analysis for wavelength interrogation method showing from (a) Reflectance VS Wavelength curves for 0.01 RIU change, (b) Reflectance VS Wavelength curves for 0.06 RIU change that small change cannot be easily detected

# STUDY OF PHOSPHOLIPID USING NARROW GROOVE PLASMONIC NANO GRATINGS

Plasmonic biosensors based on basic kretschmann configuration do not show enough sensitivity to detect very small change in lipid layer in wavelength interrogation technique, but narrow groove metallic nano grating structure [28] can detect very small changes of lipid for wide range of wavelengths. The simulated structure is given in Fig. 6, where we varied the groove width and height keeping the groove periodicity fixed at 62 nm and lipid layer thickness at 30 nm. For this case, the metallic nano gratings are filled with tryptophan protein as ligand. Then lipid solution is passed over it which is shown by lipid layer.

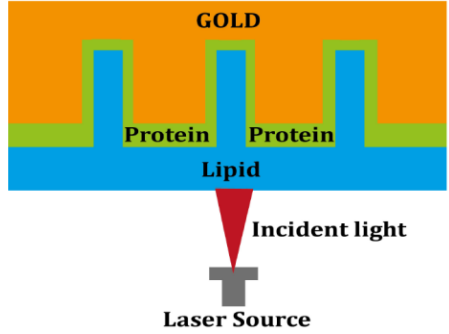

Fig. 6. Narrow Groove Plasmonic Nano Grating

The results we observed are shown in Fig. 7 and tabulated in Table IV from which we calculated various performance parameters. The groove width is varied keeping the height fixed, and again height is varied keeping the width fixed. There is significant amount of change in the curve with respect to height and width. For a grating stucture of fixed width and height the shift in resonance wavelength can be observed with respect to slight change in refractive index of phospholipid. Sensitivity (nm/RIU) was calculated for different configuration by this way.

TABLE I
SIMULATION WITH GOLD, TRYPTOPHAN AND PHOSPHOLIPID

| Gold Thickness (nm) | Lipid Thickness (nm) | SPR angle without Lipid | SPR angle with Lipid | Difference between SPR angle | Minimum reflectivity | FWHM | Sensitivity $(d\theta/d\eta)$ | Figure of Merit |
|---|---|---|---|---|---|---|---|---|
| 50 | 10 | 47.07 | 49.03 | 1.96 | 0.0052 | 3 | 3.42 | 219.23 |
|  | 20 |  | 51.38 | 4.31 | 0.0054 | 5 | 10.44 | 386.67 |
|  | 30 |  | 54.38 | 7.31 | 0.0056 | 6 | 17.54 | 522 |
| 45 | 10 | 47.24 | 49.14 | 1.90 | 0.0493 | 3.92 | 3.44 | 17.80 |
|  | 20 |  | 51.55 | 4.31 | 0.0514 | 5.28 | 10.34 | 38.09 |
|  | 30 |  | 54.21 | 6.97 | 0.0551 | 6.50 | 21.95 | 61.28 |
| 40 | 10 | 47.41 | 49.34 | 1.93 | 0.1336 | 3.98 | 3.45 | 6.48 |
|  | 20 |  | 51.76 | 4.35 | 0.1387 | 5.22 | 8.62 | 11.91 |
|  | 30 |  | 54.95 | 7.54 | 0.1452 | 6.20 | 21.06 | 23.39 |

TABLE II
SIMULATION WITH GOLD, BSA AND PHOSPHOLIPID

| Gold Thickness (nm) | Lipid Thickness (nm) | SPR angle without Lipid | SPR angle with Lipid | Difference between SPR angle | Minimum reflectivity | FWHM | Sensitivity $(d\theta/d\eta)$ | Figure of Merit |
|---|---|---|---|---|---|---|---|---|
| 50 | 10 | 47.05 | 48.69 | 1.64 | 0.0044 | 3.11 | 3.45 | 252.12 |
|  | 20 |  | 51.21 | 4.16 | 0.0051 | 4.88 | 8.62 | 346.35 |
|  | 30 |  | 54.16 | 7.11 | 0.0055 | 6.35 | 16.66 | 477 |
| 45 | 10 | 47.37 | 48.96 | 1.59 | 0.0484 | 3.91 | 2.05 | 10.83 |
|  | 20 |  | 50.82 | 3.45 | 0.0501 | 4.6 | 9.32 | 40.44 |
|  | 30 |  | 54.26 | 6.89 | 0.0537 | 6.86 | 17.24 | 46.79 |
| 40 | 10 | 47.16 | 49.26 | 2.1 | 0.1322 | 3.77 | 5.26 | 10.55 |
|  | 20 |  | 51.58 | 4.21 | 0.1377 | 5.13 | 8.77 | 12.41 |
|  | 30 |  | 54.68 | 7.52 | 0.1438 | 6.4 | 20.68 | 22.47 |

For a fixed groove width, wavelength for exciting SPR increases with groove height, whereas the minimum reflectance decreases. If the width is increased and the height is decreased resonance condition occurs at a very large wavelength as can be seen from the results in Table IV. Minimum reflectivity is also very high for lower height, so, detection of any change will be difficult. Thus, smaller width and larger height of the groove are required for biosensing Lipid bound with protein. Moreover, sensitivity is very high for nano gratings and it increases with groove height for a fixed period and width. The highest sensitivity (900 nm/RIU) is found at groove width of 10 nm and height of 50 nm for 1411 nm wavelength. It is also found that increasing height is better for any configuration because it shows lowest reflectivity. Our data also show that the wavelengths for resonance condition are in the infrared region.

TABLE III
SIMULATION WITH GOLD, TRYPTOPHAN AND EGGYOLK

| Gold Thickness (nm) | Lipid Thickness (nm) | SPR angle without Lipid (degree) | SPR angle with Lipid (degree) | Difference between SPR angle (degree) | Minimum reflectivity | FWHM (degree) | Sensitivity ($d\theta/d\eta$) | Figure of Merit |
|---|---|---|---|---|---|---|---|---|
| 50 | 10 | 47.07 | 48.83 | 1.76 | 0.005564 | 3.11 | 3.45 | 199.37 |
|  | 20 |  | 50.76 | 3.69 | 0.005589 | 4.48 | 10.34 | 412.96 |
|  | 30 |  | 53.26 | 6.19 | 0.005660 | 6.1 | 17 | 492.38 |
| 45 | 10 | 47.24 | 48.93 | 1.69 | 0.04888 | 3.62 | 3.45 | 19.49 |
|  | 20 |  | 50.97 | 3.73 | 0.05077 | 4.99 | 10.25 | 40.46 |
|  | 30 |  | 53.45 | 6.21 | 0.05305 | 6.29 | 17.24 | 51.66 |
| 40 | 10 | 47.41 | 49.13 | 1.72 | 0.1332 | 3.79 | 3.44 | 6.81 |
|  | 20 |  | 51.24 | 3.83 | 0.1375 | 4.88 | 10.34 | 15.41 |
|  | 30 |  | 53.74 | 6.33 | 0.1440 | 6.12 | 17.09 | 19.39 |

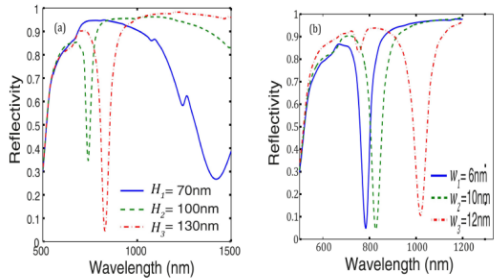

Fig. 7. (a) Reflectance with respect to wavelength for different groove height ($H$), (b) Reflectance with respect to wavelength for different groove width ($W$)

TABLE IV
DATA FOR NARROW GROOVE NANO GRATING WITH LIPID AS SENSING LAYER

| Groove Width(nm) | Groove Height (nm) | Resonance Wavelength (nm) | Minimum Reflectivity | Sensitivity (nm/RIU) |
|---|---|---|---|---|
| 6 | 70 | 753.6 | 0.7781 | 200 |
|  | 100 | 869.7 | 0.2944 | 315 |
|  | 130 | 1015 | 0.1042 | 400 |
| 10 | 70 | 1411 | 0.2713 | 900 |
|  | 100 | 750.9 | 0.01731 | 295 |
|  | 130 | 825 | 0.01731 | 355 |
| 12 | 70 | 1312 | 0.03531 | 800 |
|  | 100 | 710.6 | 0.4473 | 185 |
|  | 130 | 779.3 | 0.01466 | 295 |

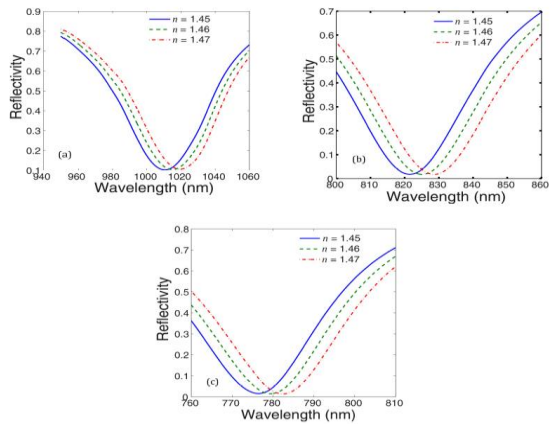

Fig. 8. Sensitivity and resonance wavelength analysis with respect to the change in refractive index($n$) for (a) 6nm Groove width and 130nm Groove height, (b) 10nm Groove width and 130nm Groove height, (c) 12nm Groove width and 130nm Groove height

# CONCLUSION

In this paper, some performance parameters such as sensitivity, minimum reflectance, FWHM and FOM for two types of SPR biosensor configurations known as basic krestchmann configuration and narrow groove nano grating coupler are analyzed with different combinations of Lipid and protein, where protein layer is used as ligand and lipid layer is considered as analyte. One of the significant findings is that the combination of phospholipid and tryptophan in the basic kretschmann configuration shows the highest sensitivity (21.95$^0$/RIU) when metal and lipid layer thicknesses are 45 nm and 30 nm respectively, however, the FOM is maximum (522) when metal layer thickness is 50 nm because the minimum reflectance is the lowest at 50 nm metal thickness. Moreover, FWHM depends on lipid layer thickness more than metal thickness. In case of the grating coupler, the variation of width and height of the grooves caused differences in the performance parameters. The periodicity of the grating were fixed at 62 nm and groove width of 10 nm and height of 70 nm gave the maximum sensitivity of 900nm/RIU in infrared region for phospholipid and tryptophan combination.